\documentclass[conference]{IEEEtran}
\usepackage{graphicx}
\usepackage{balance}
\usepackage{amsmath}
\usepackage{amssymb}
\usepackage{multicol}

\usepackage{hyperref}
\usepackage{epsf}
\usepackage{acronym}

\acrodef{CPFSK}[CPFSK]{Continuous-phase frequency-shift keying} 
\acrodef{FH}[FH]{Frequency-hopping}   
\acrodef{ACI}[ACI]{adjacent-channel interference}
\acrodef{CCI}[CCI]{co-channel interference}
\acrodef{DS-CDMA}[DS-CDMA]{direct-sequence code-division-multiple-access}
\acrodef{OFDM}[OFDM]{orthogonal-frequency-division-multiplexing}
\acrodef{LTE}[LTE]{long-term-evolution}
\acrodef{SINR}[SINR]{signal-to-interference-and-noise ratio}
\acrodef{MCS}[MCS]{modulation and coding scheme}
\acrodef{AWGN}[AWGN]{additive white Gaussian noise}
\acrodef{TC}[TC]{transmission capacity}
\acrodef{MASE}[MASE]{modulation-constrained area spectral efficiency}
\acrodef{SNR}[SNR]{signal-to-noise ratio}
\acrodef{BPP}[BPP]{binomial point process}
\acrodef{CDF}[CDF]{cumulative distribution function}
\acrodef{LOS}[LOS]{line-of-sight}
\acrodef{DS-CDMA}[DS-CDMA]{direct-sequence code-division multiple-access}
\acrodef{OFDMA}[OFDMA]{orthogonal frequency-division multiple-access}

\usepackage[usenames]{color}

\begin{document}
\hyphenation{multi-symbol}
\title{Optimization of an Adaptive \\ Frequency-Hopping Network}
\author{\IEEEauthorblockN{ Salvatore Talarico,\IEEEauthorrefmark{1} Matthew C. Valenti\IEEEauthorrefmark{1} and
Don Torrieri,\IEEEauthorrefmark{2}}
\IEEEauthorblockA{\IEEEauthorrefmark{1}West Virginia University, Morgantown, WV, USA. \\
\IEEEauthorrefmark{2}U.S. Army Research Laboratory, Adelphi, MD, USA.}
}
\date{}
\maketitle

\vspace{-0.8cm}
\thispagestyle{empty}

\begin{abstract}
This paper proposes a methodology for optimizing a  frequency-hopping network that uses continuous-phase frequency-shift keying and adaptive capacity-approaching channel coding. The optimization takes into account the spatial distribution of the interfering mobiles, Nakagami fading, and lognormal shadowing.  It includes the effects of both \acl{CCI} and \acl{ACI}, which arises due to spectral-splatter effects. The average network performance depends on the choice of the modulation index, the number of frequency-hopping channels, and the fractional in-band power, which are assumed to be fixed network parameters.  The performance of a given transmission depends on the code rate, which is adapted in response to the  expected interference to meet a constraint on outage probability.  The optimization proceeds by choosing a set of fixed network parameters, drawing the interferers from the spatial distribution, and determining the maximum rate that satisfies the outage constraint.  The process is repeated for a large number of network realizations, and the fixed network parameters that maximize the area spectral efficiency are identified.
\end{abstract}

\section{Introduction} \label{Section:Intro}

\ac{FH} spread spectrum is a versatile channel access strategy for ad hoc networks \cite{torrieri:2015}. With \ac{FH}, a transmitter and its receiver periodically hop within a large set of frequency channels over a given bandwidth according to a pseudo-random sequence.  When the transmitter and the receiver remain tuned to the same frequency, they are able to communicate.  \ac{FH} can exploit a much wider bandwidth than is possible with a single carrier frequency, and the band can be divided into noncontiguous segments. Furthermore, \ac{FH} provides inherent resistance to the near-far problem without requiring power control.



\ac{CPFSK} is a commonly used modulation for \ac{FH} systems, since it has a very compact spectrum and produces little spectral splatter.   \ac{CPFSK} modulation is characterized by its modulation order, i.e., the number of possible tones, and by its modulation index  $h$,  which  is  the  normalized  tone  spacing.  For  a  fixed modulation order, the selection of $h$ involves a tradeoff between bandwidth and performance. The bandwidth efficiency generally increases with decreasing $h$, while the error rate generally decreases with increasing $h$. In an ad hoc network, the tradeoff can be quantified by establishing a threshold on the \ac{SINR}, which quantifies the minimum signal quality required for reliable communication.  When a capacity-approaching code is used, there is a direct connection between the \ac{SINR} threshold and the achievable rate.  Together, the choice of modulation order, modulation index, and code rate constitute the {\em \ac{MCS}} for the system.

In \cite{cheng:ciss2007}, the capacity of noncoherently detected \ac{CPFSK} is computed in the absence of interference in both \ac{AWGN} and Rayleigh fading channels.   Leveraging these results, \cite{torrieri:2008} optimizes a \ac{FH} system in the presence of partial-band jamming and multiple-access interference by constraining the bandwidth, selecting the modulation index and code rate that is optimal for a Rayleigh fading channel, and fixing the number of frequency channels according to the bandwidth constraint and \ac{MCS} parameters.  The optimization is improved in \cite{valenti:2012} by jointly optimizing the \ac{MCS} parameters and the number of frequency channels when the interferers are drawn from an arbitrary spatial distribution and the channel is subject to Nakagami fading and lognormal shadowing.

The optimization of \cite{valenti:2012} leverages the analysis of \cite{torrieri:2012}, which computes the outage probability conditioned on the network geometry, then averages the outage probability with respect to the interferers' spatial distribution.  The optimization is with respect to the {\em transmission capacity} \cite{weber:2005,weber:2010}, a measure of area spectral efficiency, which is adapted to account for the modulation constraints; here, we refer to it as {\em \ac{MASE}}. While \cite{torrieri:2008,valenti:2012} only consider the influence of \ac{CCI}, \ac{ACI} is also of concern. In \cite{valenti:2013icc} the analysis and optimization is extended to account for \ac{ACI} with the goal of determining the proper amount of spectral splatter, which may be quantified by the \emph{fractional in-band power}; i.e., the amount of signal energy contained within the hopping band.  Contrary to the traditional use of a 99\% fractional in-band power, it is found that reducing the in-band power to 95\% improves throughput despite increasing \ac{ACI}, due to allowing for either more frequency channels, which reduces the likelihood of a collision, or supporting a lower-rate code, which provides more error protection when a collision does occur.

In all of these previous works, a globally optimal set of parameters is found for the network.  The parameters include the modulation index, number of frequency channels, fractional in-band power, and code rate.  However, by fixing these parameters, the outage probability will vary from one network realization to the next.  Thus, as interferers move, join, or leave the network, there is a potential for wide swings in the outage probability, which results in an unacceptably large delay jitter when a retransmission protocol is used.  For this reason, it is desirable to adapt the system in accordance with the particular network realization in an effort to meet a constraint on outage probability.  When adapting a \ac{FH} system, it is not desirable to change the number of frequency-hopping channels, the modulation index, or the fractional in-band power, as these parameters should be common to the system.  However, the rate can and should be adjusted in order to meet the performance objectives.

In this paper, we consider the impact of adapting the rate on an ad hoc \ac{FH} network.  While assuming that the rate is adapted places this work closer to what is done in practice, it has a significant impact on the analysis because special care must be taken in how the averaging is done over the spatial and fading distributions.   When the rate is fixed, the outage probability can be found by averaging over both the spatial and fading distributions.  However, when the rate is adapted, the outage probability is typically constrained to not exceed some threshold, and thus the rate must be found that meets the outage constraint for each network realization. In addition to averaging the rate over space, the distribution of rates can be determined and used as a measure of throughput and fairness.

The remainder of the paper is organized as follows.  Section \ref{Section:SystemModel} describes the system model, including a discussion of the \ac{SINR} and conditional outage probability.   Section \ref{Section:AdaptiveRate} discusses rate adaptation. Section \ref{Section:Optimization} defines \ac{MASE},  describes the procedure used for determining the combination of parameters that maximize the \ac{MASE}, and provides optimization results for several channel and network models. Finally, the paper concludes in Section \ref{Section:Conclusion}.

\section{Network Model} \label{Section:SystemModel}
Consider a finite ad hoc network composed of $M+2$ mobiles: a reference receiver, a reference source transmitter $X_{0}$, and $M$ interfering mobiles $X_{1},...,X_{M}.$   The coordinate system is chosen such that the reference receiver is at the origin.  The interfering mobiles are constrained to lie in a finite region of area $A$. The variable $X_{i}$ represents both the $i^{th}$ mobile and its location, and $||X_{i}||$ is the distance from $X_i$ to the receiver.

\subsection{SINR}

When $X_i$ transmits a signal with an omnidirectional antenna, its power at the reference receiver is
\begin{eqnarray}
  \rho_i
  & = &
  P_i g_i 10^{\xi_i/10} f( ||X_i|| ) \label{eqn:power}
\end{eqnarray}
where $P_{i}$ is the average received power from mobile $X_{i}$ at a reference distance $d_0$ when both fading and
shadowing are absent, $g_i$ is the power gain due to fading, $\xi_i$ is a shadowing coefficient, and $f( d )$ is a path-loss function. The power gain due to fading is modeled as $g_i = a_i^2$, where $a_i$ has a Nakagami distribution with parameter $m_i$, and $g_i$ has unit mean.  In Rayleigh fading, $m_i=1$ and $g_i$ is exponential.
While the $\{g_{i}\}$ are independent, they are not necessarily identically distributed.
Furthermore, the channel from each transmitting mobile to the reference receiver can have a distinct Nakagami parameter $m_i$. In the presence of lognormal shadowing, the $\{ \xi_i \}$ are independent and identically distributed zero-mean Gaussian with standard deviation $\sigma_\mathsf{s}$ dB.  In the absence of shadowing, $\xi_i = 0$.  \ For $d \geq d_0$, the path-loss function is expressed as the attenuation power law
\begin{eqnarray}
   f \left( d \right)
   & = &
   \left( \frac{d}{d_0} \right)^{-\alpha} \label{eqn:pathloss}
\end{eqnarray}
where $\alpha > 2$ is the attenuation power-law exponent, and $d_0$ is sufficiently large that the signals are in the far field.

Channel access is through a frequency-hopping protocol.  The hopping is slow, with multiple symbols per hop, which is a more suitable strategy for ad hoc networks than fast hopping \cite{torrieri:2015}.  An overall spectral band of $B$ Hz is divided into $L$ contiguous frequency channels, each of bandwidth $B/L$ Hz. Let $D_i \leq 1$ be the duty factor of $X_i$, which is the probability that the mobile transmits any signal, here assumed to be the same for each mobile ($D_i = D,\forall i$).
The system is characterized by two types of interference: a) {\em \ac{CCI}}, which is when the source and the interfering mobile use the same frequency and occurs with probability $p_\mathsf{c}$, and b) {\em \ac{ACI}}, which is when the source and the interfering mobile select adjacent channels and occurs with probability $p_\mathsf{a}$.

Each $X_i$ independently selects its frequency channel with probability $1/L$ and transmits with probability $D$.  Therefore, the probability of \ac{CCI} is $p_\mathsf{c} = D/L$. The frequency channel at the edge of the band is selected by the source $X_0$ with probability $2/L$, while an interior channel is chosen with probability $(L-2)/L$. A mobile $X_i$, $i\geq 0$, transmits by using a frequency channel adjacent to the one selected by the source $X_0$ with probability $1/L$ if $X_0$ selected a frequency channel at the edge of the band, where there is only one adjacent channel; otherwise, the mobile selects an adjacent channel of the source with probability $2/L$.
It follows that the probability that $X_i$ produces \ac{ACI} is
\begin{eqnarray}
p_\mathsf{a} =  D\left[ \left( \frac{2}{L} \right) \left( \frac{L-2}{L} \right) + \left( \frac{1}{L} \right) \left( \frac{2}{L} \right) \right] = \frac{2D(L-1)}{L^2}.\label{eqn:p_a}
\end{eqnarray}

When $X_i$ transmits, only a fraction of the total transmitted power lies within the occupied frequency channel.  This fraction is called {\em fractional in-band power}, and is denoted by $\psi$. Typically, $0.95 \leq \psi \leq 0.99$. The remaining fraction of the power is assumed to spill significantly only into the frequency channels that are adjacent to the one selected, which is a reasonable assumption for most practical systems. Let $K_s = (1-\psi)/2$ be the {\em adjacent-channel splatter ratio}, which indicates the fraction of power that spills into each adjacent channel.

By using a {\em physical interference model} \cite{cardieri:2010}, the instantaneous \ac{SINR} at the reference receiver is
\begin{eqnarray}
   \gamma
   & = &
   \frac{ \psi g_0 \Omega_0  }{ \displaystyle \mathsf{SNR}^{-1} + \sum_{i=1}^M I_i g_i \Omega_i }
   \label{Equation:SINR2}
\end{eqnarray}
where $\mathsf{SNR} = d_0^\alpha P_{0}/\mathcal{N}$ is the \ac{SNR}  when the transmitter is at unit distance and fading and
shadowing are absent, $\mathcal N$ is the noise power,
\begin{equation}
\Omega_{i}=%
\begin{cases}
10^{\xi_{0}/10}||X_{0}||^{-\alpha} & i=0 \\
\displaystyle\frac{P_{i}}{P_{0}}10^{\xi_{i}/10}||X_{i}||^{-\alpha} & i\geq1%
\end{cases}
\label{eqn:omega}
\end{equation}
is the normalized received power due to $X_{i}$ and $I_i$ is a discrete random variable that may take on three values with the probabilities
\begin{eqnarray}
I_i
& = &
\begin{cases}
  \psi & \mbox{with probability $p_\mathsf{c}$} \\
  K_s & \mbox{with probability $p_\mathsf{a}$} \\
  0 & \mbox{ with probability $p_\mathsf{n}$ }
\end{cases}
\label{eqn:I_i}
\end{eqnarray}
where $p_\mathsf{n} = 1 - p_\mathsf{c} - p_\mathsf{a} = 1-D(3L-2)/L^2$ is the probability of no collision. The three probabilities correspond to the case that $X_i$ selects the same frequency channel as $X_0$, $X_i$ selects a frequency channel that is adjacent to the one chosen by $X_0$, or $X_i$ selects a distant frequency channel (more than two channels from the one chosen by $X_0$). \ac{ACI} can be neglected by setting $\psi=1$ and $K_s=0$, in which case the model specializes to the one in \cite{valenti:2012}.

\subsection{Conditional Outage Probability}\label{Section:Outage}

Let $\beta$ denote the minimum \ac{SINR} required for reliable reception and $\boldsymbol{\Omega }=\{\Omega_{0},...,\Omega _{M}\}$ represent the set of normalized received powers.  An \emph{outage} occurs when the \ac{SINR} falls below $\beta$.  When conditioned on $\boldsymbol{\Omega }$, the outage probability is
\begin{eqnarray}
   \epsilon( \boldsymbol{\Omega } )
   & = &
   P \left[ \gamma \leq \beta \big| \boldsymbol \Omega \right]\hspace{-0.1cm}.
   \label{Equation:Outage1}
\end{eqnarray}
Assuming that the Nakagami parameter $m_0$ of the source's channel is integer-valued and under the assumption of block fading, the analysis of \cite{torrieri:2012} using (\ref{eqn:I_i}) gives
\begin{eqnarray}
\epsilon( \boldsymbol{\Omega } )
=
\exp\left(-\frac{\beta_{0}}{\mathsf{SNR}}\right)\sum_{s=0}^{m_{0}-1}{\left( \frac{\beta_{0}}{\mathsf{SNR}}\right) }%
^{s}\sum_{t=0}^{s}\frac{H_{t}( \boldsymbol{\Omega})}{\mathsf{SNR}^{t}(s-t)!}
\label{Equation:NakagamiConditional}
\end{eqnarray}
where $\beta_0 = \beta m_0 / \left(\psi \Omega_0\right) $, and
\begin{eqnarray}
H_{t}(\boldsymbol{\Omega})
& = &
\mathop{ \sum_{\ell_i \geq 0}}%
_{\sum_{i=1}^{M}\ell_{i}=t}\prod_{i=1}^{M}G_{\ell_{i}}(\Omega_{i}).
\label{Equation:Hfunc}
\end{eqnarray}
The summation in (\ref{Equation:Hfunc}) is over all sets of positive indices that sum to $t$, and from \cite{valenti:2013icc}
\begin{eqnarray}
  G_{\ell_i}( \Omega_i )
  =
p_\mathsf{n} \delta_{\ell_i}
  +
    \frac{ \Gamma(\ell_i+m_i) }{ \Gamma(\ell_i+1) \Gamma( m_i ) }
\left[ p_\mathsf{c} \phi_i( \psi )
+
  p_\mathsf{a}
  \phi_i( K_s )
 \right]
 \label{Equation:Gfunc}
\end{eqnarray}
where $\delta_{\ell}$ is the Kronecker delta function, equal to $1$ when $\ell=0$, and zero otherwise, $\Gamma\left( x\right)$ is the Gamma function,
 and
\begin{eqnarray}
  \phi_i( x )
  & = &
  \left( \frac{x \Omega_i}{ m_i} \right)^{\ell_i}
  \left(
   \frac{ x \beta_0 \Omega_i}{m_i} + 1
 \right)^{-(m_i+\ell_i)}.
\end{eqnarray}
The conditional outage probability is calculated once the location of the mobiles and the shadowing factors are specified. Thus, performance can be calculated for any snapshot of the network topology.

\section{Rate Adaptation}\label{Section:AdaptiveRate}
In \ac{FH} systems, the probability of a collision can be reduced by increasing the number of frequency channels, and therefore it is highly desirable to choose a data modulation that has a compact spectrum and a reduced spectral splatter.  Because \ac{CPFSK} has these characteristics \cite{torrieri:2015}, it is the preferred modulation for \ac{FH} systems.   When used for \ac{FH} applications, \ac{CPFSK} is typically detected noncoherently since rapid hopping precludes carrier synchronization.  A \ac{CPFSK} modulation is characterized by its modulation order $q$, which is the number of possible tones, and by its modulation index $h$, which is the normalized tone spacing. For \ac{FH} networks, performance also depends on the number of frequency channels, the rate $R$ of the error-control code,  and the fractional in-band power.

The rate is often adapted (e.g., \cite{wang:2002,Lee:2013}) by assigning a different code rate based on the quality of the channel to increase the system capacity and/or to satisfy a specific outage constraint.  While in \cite{valenti:2013icc}, an optimal common rate is found together with the modulation characteristics in order to maximize the throughput of a \ac{FH} system, in this work the effect of link adaptation is taken into account by assuming that a mobile is able to estimate the channel and appropriately adapt its rate in order to meet a given outage constraint.  The adaptation assumes that the distribution of the outage probability is known; i.e., that the transmitter has some way to estimate (\ref{Equation:Outage1}) for its intended receiver.  This function can be precisely estimated if the transmitter locations, transmitter powers, shadowing coefficients, and the parameters in (\ref{eqn:I_i}), i.e. $\psi$, $K_s$, $p_\mathsf{c}$ and $p_\mathsf{a}$, are known.  Importantly, the instantaneous fading gains or the particular channel selected by the transmitter need not be known.  In the absence of these parameters, a simple adaptive scheme could request a higher rate when the observed outage probability (as determined by a CRC check) is below the target and requests a lower rate when it is above.

Let $C(\gamma)$ be the capacity as a function of the SINR $\gamma$.  The channel is in an outage when $C(\gamma) \leq R$.  Assuming Gaussian interference, the channel is conditionally Gaussian during a particular hop and the modulation-constrained \ac{AWGN} capacity can be used for $C(\gamma)$. The maximum achievable rate of noncoherent \ac{CPFSK} is given in \cite{cheng:ciss2007} for various modulation indices $h$, where it is called the {\em symmetric information rate}. In particular, Fig. 1 of \cite{cheng:ciss2007} shows the symmetric information rate of binary \ac{CPFSK} as a function of $\gamma$ for various $h$.  To emphasize the dependency of the capacity on $h$, $C(h,\gamma)$ is used in the sequel to denote the rate of \ac{CPFSK} with modulation index $h$. For any value of $h$ and $\gamma$, the achievable code rate is $R=C(h,\gamma)$ for the given $\gamma = \beta$. When link adaptation is used, the \ac{SINR} threshold $\beta$ is selected such that $P \left[\gamma \leq \beta \right]<\hat\epsilon$,  where $\hat\epsilon$ is a constraint on the channel outage. The appropriate $\beta$ varies according to the number of frequency channels available and the fractional in-band power chosen.

\section{Network Optimization}\label{Section:Optimization}

For an ad hoc \ac{FH} network that uses \ac{CPFSK} modulation and adaptive-rate coding, there will be an optimal set of $(L,h,\psi)$ for the particular spatial distribution.   The optimization is performed with respect to the \ac{MASE}, which is defined in subsection \ref{Section:TC}.  The optimization algorithm is presented in subsection \ref{Section:Algorithm}, and the results are given in subsection \ref{Section:OptimizationResults}.

\subsection{Objective}\label{Section:TC}
The transmission capacity \cite{weber:2005,weber:2010} is a measure of the area spectral efficiency; i.e., the maximum rate of successful data transmission per unit area.  The transmission capacity may be found by multiplying the average throughput of a typical link by the density of transmitters
\begin{eqnarray}
\tau\left(\lambda\right)=\lambda\mathbb{E}\left[T \right]
\label{TC_definition}
\end{eqnarray}
where $\lambda = M/A$ and $\mathbb{E}\left[T \right]$ is the throughput averaged over the network topologies.  When accounting for modulation and coding, the maximum data transmission rate is determined by the bandwidth $B/L$ of a frequency channel, the spectral efficiency of the modulation, and the code rate.

Let $\eta$ denote the spectral efficiency of the modulation, which is defined by the symbol rate divided by the $100\psi$\%-power bandwidth of the modulation.   Neglecting the switching effects of finite-duration hops, the spectral efficiency of \ac{CPFSK} can be found by numerically integrating the normalized power-spectral density given by Equation (3.4-61) of \cite{proakis:2008} and then inverting the result.  To emphasize the dependence of $\eta$ on $h$ and $\psi$, we denote the spectral efficiency of \ac{CPFSK} as $\eta\left(h,\psi\right)$.  When combined with a rate-$R$ code, the spectral efficiency of \ac{CPFSK} becomes $R \eta\left(h,\psi\right)$ (information) bits per second per Hz.  The data rate supported by the channel is $R \eta\left(h,\psi\right) B/L$ bits per second.  The throughput must account for the duty factor $D$ and only count correct transmissions.  Hence, the throughput is
\begin{eqnarray}
   T
   & = &
   \frac {   R \eta(h,\psi) B D (1-\hat\epsilon) }{L} \label{eq:throughput}
\end{eqnarray}

By substituting (\ref{eq:throughput}) into (\ref{TC_definition}), and dividing by bandwidth $B$, the normalized \ac{MASE} is
\begin{eqnarray}
   \tau' (\lambda)
   & = & \frac{\tau}{B} = \frac{\lambda \mathbb{E}\left[R\right] \eta\left(h,\psi\right) D (1-\hat\epsilon) }{L}.\label{Equation:TC}
\end{eqnarray}
In contrast with (\ref{TC_definition}), this form of transmission capacity explicitly takes into account the code rate $R$, as well as the spectral efficiency of the modulation $\eta\left(h,\psi\right)$ and the equivalent number of frequency channels.

In the literature, the transmission capacity is typically interpreted to be the spatial density of transmissions that achieves a target \emph{average} outage probability \cite{weber:2010}.  In contrast, we place a constraint on the \emph{conditional} outage probability; i.e., the outage probability of a particular network realization for which we are adapting the rate.  Thus, the MASE developed in this section can be interpreted as the spatial density of transmission that achieves a target \emph{conditional} outage probability, where the conditioning is with respect to the network topology.

\subsection{Optimization Algorithm}\label{Section:Algorithm}

For the optimization, the density of potentially interfering mobiles per unit area $\lambda$ and duty factor $D$ have fixed values. For a given modulation index $h$, a given number of frequency channels, and a given fractional in-band power, the average code rate $\mathbb{E}\left[R\right]$ can be found using a Monte Carlo approach as follows. Draw a realization of the network by placing $M$ mobiles within an area $A$ according to a specific spatial distribution. Compute the path loss from each mobile to each reference receiver, applying randomly generated shadowing factors. By  setting  the  outage equal to the outage constraint $\hat\epsilon$, invert the outage probability expression given by (\ref{Equation:NakagamiConditional}) to determine the adequate \ac{SINR} threshold $\beta$. Next, find the rate $R_n= C(h,\beta)$ that corresponds to the \ac{SINR} threshold for the $n^{th}$ topology and store its value. Repeat this process for a large number of networks $N$. Compute the average code rate as follows
\begin{eqnarray}
   \mathbb{E}\left[R\right]
   & = & \frac{1}{N} \sum_{n=1}^N R_{n}. \label{eq:AverageRate}
\end{eqnarray}

The optimization can be accomplished using an exhaustive search by performing the following steps:
\begin{enumerate}
\item Pick a value of $h$. \label{step1}
\item Pick a value of $\psi$. \label{step2}
\item Pick a value of $L$. \label{step3}
\item Compute the average code rate  $\mathbb{E}\left[R\right]$, as described above, and find the spectral efficiency $\eta(h,\psi)$ corresponding to the current $h$ and $\psi$.
\item Use (\ref{Equation:TC}) to compute the normalized \ac{MASE}.
\item Return to step \ref{step3} until all $L$ are considered.
\item Return to step \ref{step2} until all $\psi$ are considered.
\item Return to step \ref{step1} until all $h$ are considered.
\end{enumerate}

The above procedure will find the $\tau'(\lambda)$ for each $(L,h,\psi)$ considered, and the optimal value of these parameters are the ones that maximize $\tau'(\lambda)$.
By limiting $L$ to be integer valued, the number of values is finite and an exhaustive search up to some maximum value is feasible.  For the exhaustive search results presented in this section, $h$ was quantized to a spacing of 0.01 over the range $0 \leq h \leq 1$ and $\psi$ was quantized to a spacing of 0.005 over the range $0.90 \leq \psi \leq 0.99$.

\subsection{Optimization Results}\label{Section:OptimizationResults}


As an example, consider a network with interfering mobiles drawn from a \ac{BPP} within an annular region. The region has inner radius $r_{\mathsf{ex}}=0.25$ and two outer radii are considered: $r_{\mathsf{net}}=2$ and $r_{\mathsf{net}}=4$. If not otherwise stated, the number of interfering mobiles is $M = 50$. The optimization is done by considering $N=$10,000 different network topologies. The path-loss exponent is $\alpha = 3$, the  duty  factor  is  $D  = 1$, the  source  is  located  at unit distance away from the receiver, the SNR is $ \mathsf{SNR}= 10$ dB, and when a shadowed scenario is examined, $\sigma_s= 8$ dB.  Binary \ac{CPFSK} is used, and the rate is adapted in order to satisfy a typical outage constraint of $\hat\epsilon=0.1$. Three fading models are considered: Rayleigh fading ($m_i = 1$ for all i), Nakagami fading ($m_i = 4$ for all i), and mixed fading ($m_0 = 4$ and $m_i = 1$ for $i \geq 1$).

\begin{figure}[t]
\centering
\hspace{-0.5cm}
\includegraphics[width=9.25cm]{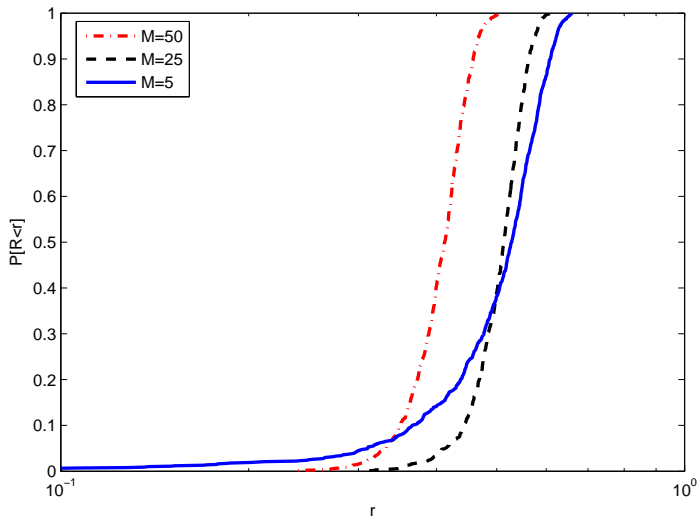}
\vspace{-0.7cm}
\caption{ \ac{CDF} of the code rate when the network is optimized. Three curves are provided: a) dense network ($M=50$); b) moderate dense network ($M=25$); c) sparse network ($M=5$).  \label{Figure:RateCDF} }
\vspace{0.85cm}
\end{figure}

\begin{figure}[t]
\centering
\hspace{-0.5cm}
\includegraphics[width=9.25cm]{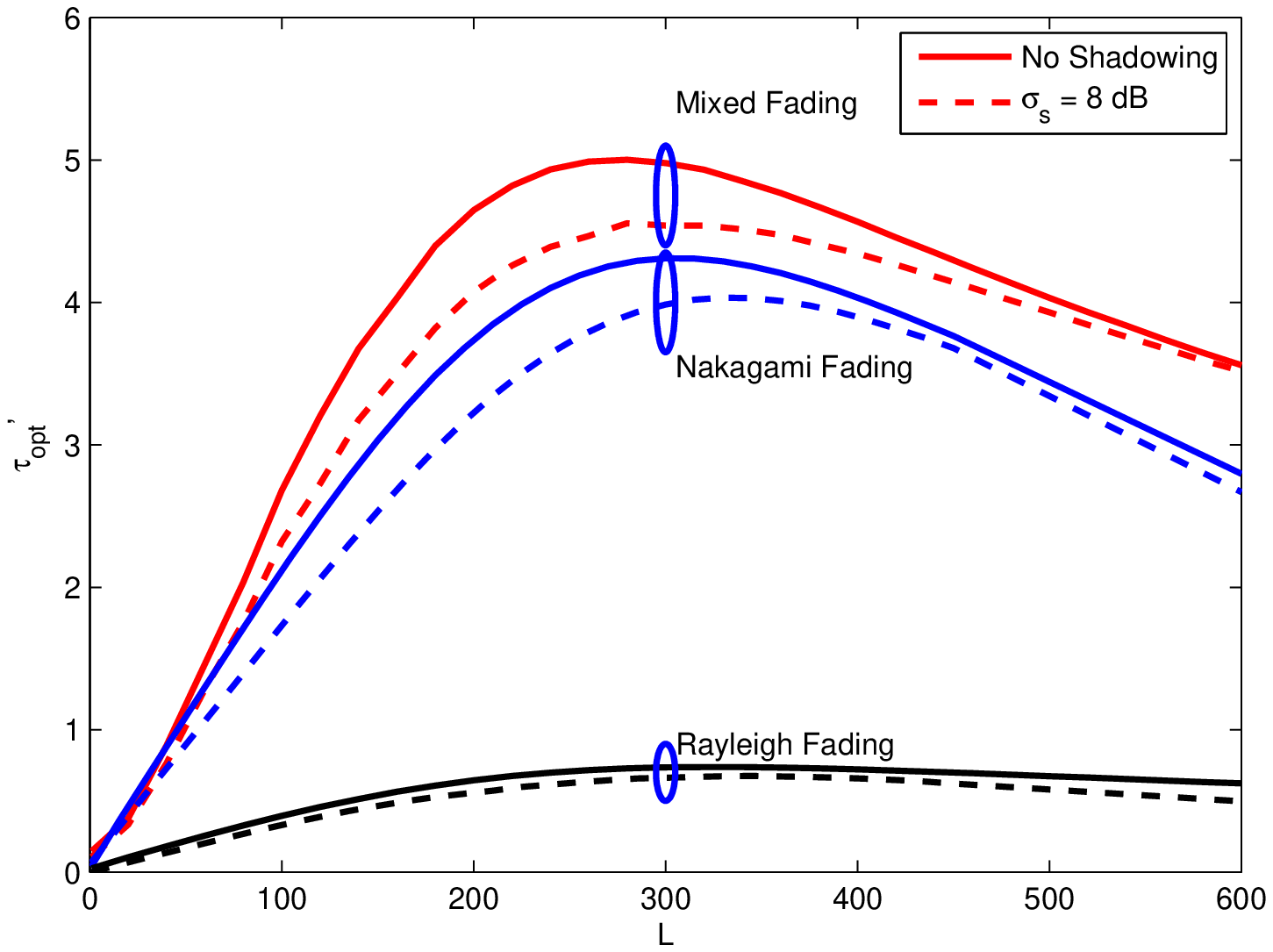}
\vspace{-0.7cm}
\caption{Optimal normalized \ac{MASE} as function of the number of frequency channels $L$. For each $L$, the modulation index is chosen in order to maximize the normalized \ac{MASE}, while $\psi=0.96$ is adopted. A dense network is considered ($M=50$ and $r_{\mathsf{net}}=2$). Top curves: mixed fading.  Middle curves: Nakagami fading.  Bottom curves: Rayleigh fading. For each channel model both a shadowed and an unshadowed scenario are considered. \label{Figure:L_TC} }
\vspace{0.25cm}
\end{figure}

\begin{figure}[t]
\centering
\hspace{-0.5cm}
\includegraphics[width=9.25cm]{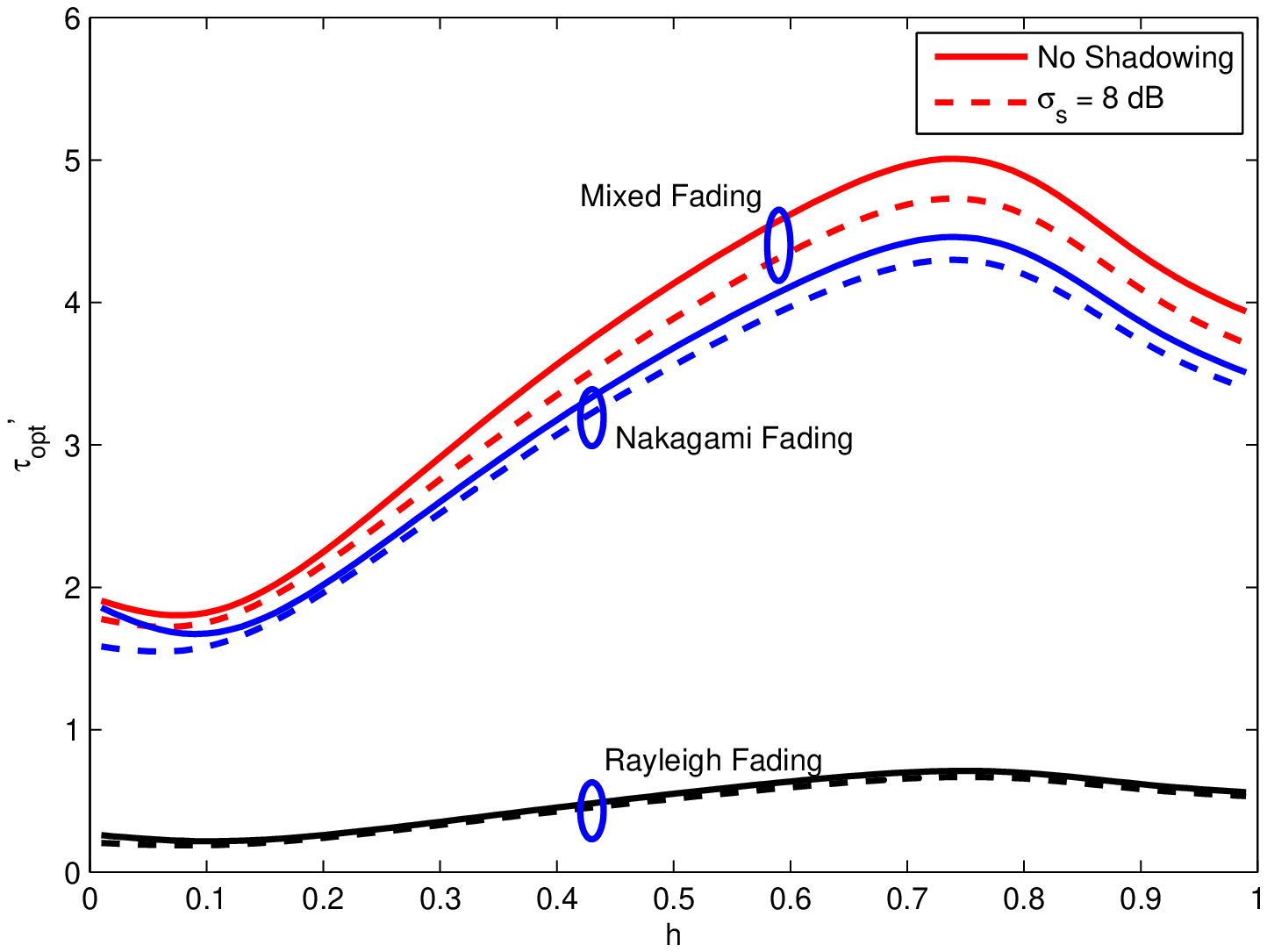}
\vspace{-0.7cm}
\caption{Optimal normalized \ac{MASE} as function of the modulation index $h$. For each $h$, the number of frequency channels is chosen in order to maximize the normalized \ac{MASE}, while $\psi=0.96$ is adopted. A dense network is considered ($M=50$ and $r_{\mathsf{net}}=2$). Top curves: mixed fading.  Middle curves: Nakagami fading.  Bottom curves: Rayleigh fading. For each channel model both a shadowed and an unshadowed scenario are considered.   \label{Figure:H_TC} }
\vspace{0.0cm}
\end{figure}

\begin{figure}[t]
\centering
\hspace{-0.5cm}
\includegraphics[width=9.25cm]{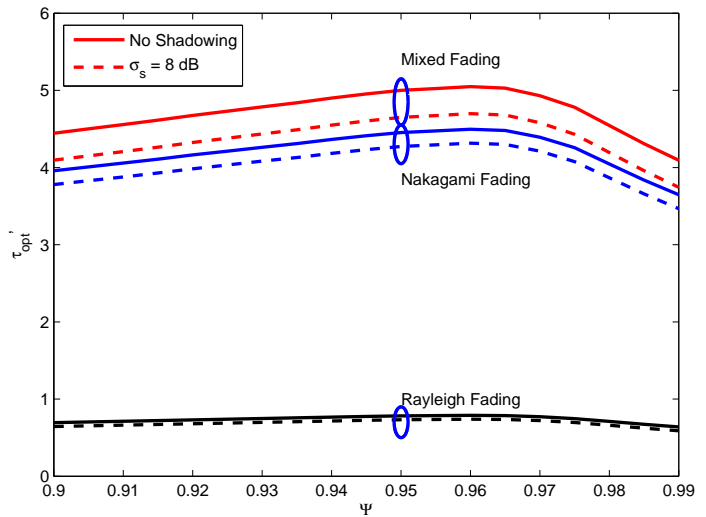}
\vspace{-0.7cm}
\caption{Optimal normalized \ac{MASE} as function of the fractional in-band power $\psi$. For each $\psi$, both the number of frequency channels and the modulation index are chosen in order to maximize the normalized \ac{MASE}. A dense network is considered ($M=50$ and $r_{\mathsf{net}}=2$). Top curves: mixed fading.  Middle curves: Nakagami fading.  Bottom curves: Rayleigh fading. For each channel model both a shadowed and an unshadowed scenario are considered.     \label{Figure:Psi_TC} }
\end{figure}

Fig. \ref{Figure:RateCDF} illustrates the variability of the code rate through its \ac{CDF} in the presence of mixed fading and shadowing. The figure shows three curves, corresponding to the case of a dense network ($M=50$), a moderately dense network ($M=25$) and a sparse network ($M=5$). The outer radius of the annular network is fixed to $r_{\mathsf{net}}=2$. The network parameters are independently optimized for each network density.  Fig. \ref{Figure:RateCDF} shows that the \ac{CDF} of the code rate is quite steep for all three scenarios, which emphasizes the fairness in the rate selection.  However, the curves are steeper for dense networks than for sparse networks, since in a sparse network there are fewer interferers and therefore more variability in the \ac{SINR} and code rate from one realization to the next.


Fig. \ref{Figure:L_TC}, \ref{Figure:H_TC}, and \ref{Figure:Psi_TC} show the optimal normalized \ac{MASE} $\tau'_{opt}$ for a dense network ($M=50$ and $r_{\mathsf{net}}=2$, respectively) as function of the number of frequency channels $L$, the modulation index $h$, and the fractional in-band power $\psi$. These three figures explore the relative importance of each of the three parameters of the optimization. For Figs. \ref{Figure:L_TC} and \ref{Figure:H_TC}, $\psi=0.96$ is used, and for each value of the given parameter ($L$ or $h$), the other one is chosen in order to maximize the normalized \ac{MASE}. For Fig. \ref{Figure:Psi_TC} the two parameters are jointly optimized for each value of $\psi$.

The results of the optimization are shown in Table \ref{maintable}.  For each set of parameters, the $(L,h,\psi)$ that maximize the normalized \ac{MASE} are listed, along with the corresponding normalized \ac{MASE} $\tau_{opt}'$.  In addition to showing the normalized \ac{MASE} when using the optimal parameters, the normalized \ac{MASE} $\tau_{sub}'$ is shown for a typical choice of parameters: $(L,h,\psi) = (200,0.5,0.99)$.

\begin{table}[t]
  \centering
  \caption{Results of the Optimization for $M=50$ interferers.  The normalized \ac{MASE} $\tau'$ is in units of bps/kHz-$m^2$. Channel abbreviations: $R$ indicates Rayleigh fading, $N$ indicates Nakagami fading ($m_i = 4$, for all $i$), $M$ indicates mixed fading ($m_0 = 4$ and $m_i=1$ for $i \geq 1$), $U$ indicates unshadowed environment, and $S$ indicates shadowing ($\sigma_\mathsf{s} = 8$ dB). \label{maintable}}
  \begin{tabular}{|c|c|c|c|c|c|c|c|}
  \hline
  $r_{\mathsf{net}}$ & Channel & $L$ & $\mathbb E[R]$ & $h$ &$\psi$& $\tau'_{opt} $ & $\tau'_{sub} $ \\
  \hline
  2         & R/U   &  315 & 0.07 & 0.80 & 0.96 & 0.79 & 0.50\\
  \cline{2-8}
            & N/U&  280 & 0.36 &  0.80 & 0.96& 4.42 & 2.99\\
  \cline{2-8}
            & M/U  & 279  & 0.41&  0.80& 0.96& 5.00 & 3.56\\
  \cline{2-8}
            & R/S   & 320 & 0.07& 0.80& 0.96 & 0.76 & 0.40\\
  \cline{2-8}
            & N/S &  290 & 0.42 & 0.80 &0.96 & 4.24 & 2.27\\
  \cline{2-8}
            & M/S & 290 &0.38 & 0.80 & 0.96& 4.70 & 3.02\\
  \hline
  4         & R/U &  73 & 0.05 & 0.84 & 0.95 &0.63 & 0.32\\
  \cline{2-8}
            & N/U&  80 & 0.35  &0.84 & 0.95& 3.74 & 1.87\\
  \cline{2-8}
            & M/U & 75 &  0.36 & 0.84 & 0.95 & 4.09 & 1.91\\
    \cline{2-8}
            & R/S & 95 &0.05 &  0.84 & 0.95  & 0.49 & 0.28\\
  \cline{2-8}
            & N/S & 130 & 0.40 &0.84 & 0.95& 2.65 & 1.75\\
  \cline{2-8}
            & M/S & 100  & 0.35 & 0.84 & 0.95& 3.03 & 1.80\\
  \hline
  \end{tabular}
\end{table}

The results shown in Table \ref{maintable} highlight the importance of parameter optimization, and point out that by selecting the optimal parameters it is possible to improve the normalized \ac{MASE} by a factor greater than two relative to arbitrary parameter selection.  Optimization results are given for different fading channel models and for both a shadowed and an unshadowed scenario. Performance is very poor for the pessimistic assumption of Rayleigh fading, while it improves for a more realistic mixed fading, which implies that the source and destination are in the \ac{LOS}, while the other links are not. Shadowing is always detrimental, and even though it leads to an higher code rate, it requires a larger number of frequency channels. While the optimal modulation index and fractional in-band power vary with the density of potentially interfering mobiles and the distance of the transmitter from the receiver, similarly to \cite{valenti:2013icc} they remain constant under different channel models and shadowing conditions.
An increment in the density of potentially interfering mobiles per unit area leads to an increment of the normalized \ac{MASE}, as well as to a higher code rate, number of frequency channels, and fractional in-band power and a lower modulation index.

\balance

\section{Conclusion} \label{Section:Conclusion}

When an \ac{FH} mobile ad hoc network uses noncoherent \ac{CPFSK} and adapts its rate in order to satisfy an outage constraint, its performance depends on the modulation index $h$, the number of frequency channels $L$, and the fractional in-band power $\psi$. A tradeoff among these parameters exists, but in practice, their values are often chosen arbitrarily. In this paper, the spatial distribution of the interfering mobiles is modeled using a point process and a physical interference model is adopted. Using a closed-form expression for the outage probability conditioned on the network geometry, the optimal code rate can be found, and by averaging over the spatial distribution, the optimal values of $h$, $L$, and $\psi$ can be found.   Using the methodology proposed here, it is shown that it is possible to improve the performance in terms of area spectral efficiency by a factor greater than two relative to an arbitrary parameter selection.

\bibliographystyle{ieeetr}
\bibliography{milcom2015refs}

\end{document}